\documentclass[twocolumn,showpacs]{revtex4}

\usepackage{graphicx}
\usepackage{dcolumn}
\usepackage{amsmath}
\begin{document}

%\twocolumn[\hsize\textwidth\columnwidth\hsize\csname
%@twocolumnfalse\endcsname
\title{Re-entrant hidden order at a metamagnetic quantum critical 
end point}
\author{N.~Harrison$^1$, M.~Jaime$^1$ and J.~A.~Mydosh$^{2,3}$
}
\address{$^1$National High Magnetic Field Laboratory, LANL, 
MS-E536, Los Alamos, New Mexico 87545\\
$^2$Kamerlingh Onnes Laboratory, Leiden University, NL-2300 RA Leiden, The 
Netherlands\\
$^3$Max-Planck Institut for Chemical Physics of Solids, N\"{o}thnitzer 
Str. 40, D-01187 Dresden, Germany
}
%\date{\today}
%\maketitle

\begin{abstract}
Magnetization measurements of URu$_2$Si$_2$ in pulsed magnetic fields
of 44~T reveal that the hidden order phase is destroyed before
appearing in the form of a re-entrant phase between $\approx$~36 and
39~T. Evidence for conventional itinerant electron metamagnetism at
higher temperatures suggests that the re-entrant phase is created in
the vicinity of a quantum critical end point.
\end{abstract}

%]\narrowtext
\maketitle

Recent studies of itinerant electron magnetism in strongly correlated
$d$- and $f$-electron metals have shown that metamagnetism gives rise
a new class of field-induced quantum phase
transition~\cite{grigera1,millis1}.  Sr$_{3}$Ru$_{2}$O$_{7}$,
CeRu$_{2}$Si$_{2}$ and UPt$_{3}$~\cite{stewart1} are all considered
examples of systems that could posses a quantum critical end point, in
which a notional line of first order phase transitions terminates at
zero rather than finite temperature~\cite{grigera1}.  Here we propose
that URu$_{2}$Si$_{2}$ may be the first example of a system in which
thermodynamic instabilities associated with the end point give rise to
an ordered phase at high magnetic fields and low
temperatures~\cite{jaime1}.  This behaviour is reminiscent of the
creation of superconductivity in the vicinity of an antiferromagnetic
quantum critical point in zero field~\cite{mathur1}.  We show that the
presence of multiple magnetic transitions in URu$_{2}$Si$_{2}$ at low
temperatures~\cite{devisser1,sakakibara1,sugiyama1} can be ascribed to
re-entrant phenomena arising from the interplay between itinerant
electron metamagnetism (IEM) and the hidden order (HO) parameter
recently attributed to orbital antiferromagnetism~\cite{chandra1}.

URu$_{2}$Si$_{2}$ belongs to a class of strongly correlated metals in
which $f$-electrons, rather than being localised and giving rise to
magnetism, develop a distinctly itinerant character~\cite{palstra1}. 
Coulomb interactions cause the quasiparticle effective masses to be
heavily renormalised, making the energetic rewards for forming ordered
groundstates substantially higher than in normal
metals~\cite{hewson1,stewart2}.  Indeed, in addition to forming the HO
phase at $T_{\mathrm{o}}\approx $~17.5~K~\cite{chandra1},
URu$_{2}$Si$_{2}$ becomes superconducting at
$T_{\mathrm{c}}\approx$~1.2~K~\cite{palstra1}.  The presence of an
itinerant $f$-electron Fermi surface \cite{ohkuni1,keller1} also
furnishes URu$_{2}$Si$_{2}$ with the essential preconditions for
IEM~\cite{edwards1}, by which the $f$-electrons revert to a localised
behaviour upon their alignment in strong magnetic fields.  IEM is
considered to account for the increase in the magnetization by
$\approx$~1~$\mu _{\mathrm{B}}$ per U atom at magnetic fields between
$\approx$~35~T and $\approx$~40~T, although the existence of multiple
magnetic transitions has remained
controversial~\cite{devisser1,sakakibara1,sugiyama1}.  Recent
observations that local moment antiferromagnetism occurs within a
minority phase that is destroyed by fields in excess of
15~T~\cite{chandra1,mason1} call for a re-examination of the bulk high
magnetic field phenomena in URu$_2$Si$_2$.

In this paper, we show that multiple transitions in URu$_{2}$Si$_{2}$
can be explained by a scenario in which the magnetic field first
destroys the HO phase before creating a new field-induced re-entrant
phase~\cite{jaime1} in the vicinity of the metamagnetic transition
(see Fig.~\ref{contour} for a phase diagram).  IEM is accompanied by a
pronounced asymmetry between the occupancy of itinerant spin-up and
spin-down $f$-electron states~\cite{edwards1}, brought on by the
sudden population of the spin-up component as it sinks below the Fermi
energy $\varepsilon _{\mathrm{F}}$ at a magnetic field $B_{\rm M}$. 
Magnetization measurements reveal that the magnetic field-induced
phase is accompanied by the opening of a gap in the spin-up
$f$-electron band at $B_{\rm M}$.  We argue that such a gap could be
compatible with a spin-singlet order parameter that breaks
translational symmetry, of which the orbital antiferromagnetic (OAF)
phase (recently proposed by Chandra \textit{et al.}~\cite{chandra1} to
explain the origin of the HO) is one such example.

Figure~\ref{raw}\textbf{a} shows examples of the magnetization $M$ of
URu$_{2}$Si$_{2}$ measured in pulsed magnetic fields of up to 44~T at
several different temperatures.  The data are obtained using a
wire-wound sample-extraction magnetometer in which the sample is
inserted or removed from the detection coils in-situ.  While the
experimental curves in Fig.~\ref{raw} appear similar to those measured
by other groups~\cite{devisser1,sakakibara1,sugiyama1}, the phase
diagram obtained in Fig.~\ref{contour} upon extracting the positions
of the maxima in the differential susceptibility $\chi=\mu_0\partial
M/\partial B|_{T}$ at different temperatures is markedly different. 
In a recent study, Jaime \textit{et al}~\cite{jaime1} noted that the
magnetocaloric effect can cause severe variations in sample
temperature in pulsed magnetic field experiments if the sample cannot
exchange heat with the bath as the magnetic field $B$ changes.  This
effect is particularly serious if the sample is too large, a poor
thermal diffusivity isolates the sample, or if the field is swept too
rapidly.  Adequate isothermal equilibrium in pulsed magnetic fields
could, however, be achieved by using a long-pulse magnet (with a field
decay constant of $\approx$~0.25~s) combined with a sample thickness
of $\simeq$~150~$\mu$m~\cite{jaime1}.  It is by making such provisions
in the present study that we obtain a phase diagram that agrees more
closely with specific heat measurements in static magnetic
fields~\cite{jaime1}.

The existence of IEM of a similar type to that observed in
Sr$_3$Ru$_2$O$_7$~\cite{perry1}, UPt$_3$~\cite{sugiyama2,frings1} and
CeRu$_2$Si$_2$~\cite{flouquet1,haen1} is evidenced at temperatures
above $\approx$~6~K in Fig.~\ref{raw}\textbf{b} by the presence of a
single broad maximum in $\chi$ .  The dashed line in
Fig.~\ref{contour} indicates that the location of this feature at
$B_{\mathrm{M}}\approx$~37.9~T does not change significantly with
temperature.  The rapid increase in the $\chi$ at $B_{\rm M}$ with
decreasing temperature, shown in Fig.~\ref{raw}\textbf{c} (filled
squares), implies that the jump in $M$ sharpens with decreasing
temperature.  Such behaviour is consistent with the existence of a
first order critical end point at a field $B_{\mathrm{M}}$ at
temperatures well below 6~K that is broadened by thermal fluctuations
at higher temperatures~\cite{millis1}.  Rather than diverging
indefinitely, however, the maximum in $\chi$ vanishes below
$\approx$~6~K on entering the field-induced ordered phase recently
indentified in specific heat measurements~\cite{jaime1}.  The fact
that $B_{\mathrm{M}}$ occurs within the field-induced ordered phase
implies that fluctuations associated with the metamagnetic critical
end-point~\cite{grigera1} could play a role in its formation.  Strong
fluctuations in the vicinity of quantum critical points can cause
metals to become highly susceptible to order as a means of lowering
energy~\cite{grigera1,millis1}.  A well known example is provided by
the creation of superconductivity in the vicinity of an
antiferromagnetic quantum critical point~\cite{mathur1}.

The intense magnetic fields combined with formation of the
field-induced phase below $\approx$~6~K in URu$_2$Si$_2$ make the
process of identifying whether the critical end point would otherwise
terminate at $T=$~0 less certain than with
Sr$_2$Ru$_3$O$_7$~\cite{grigera1}, UPt$_3$~\cite{sugiyama2,frings1} or
CeRu$_2$Si$_2$~\cite{flouquet1,haen1}.  This normally requires
evidence for non-Fermi liquid behavior.  Fortunately, the transition
in the specific heat $C$ at $\approx$~5~K~\cite{jaime1} appears to be
first order (being of $\approx$~0.25~K in width, albeit without
observable hysteresis), implying that the region above $\approx$~6~K
is free from thermal fluctuations of the field-induced HO parameter. 
This region can therefore be investigated for non-Fermi liquid effects
associated with IEM~\cite{stewart1}.  Transport studies are thus far
incomplete, presently yeilding only a broad maximum in the
magnetoresistance at $B_{\mathrm{M}}$~\cite{jaime1}.  The strongly
divergent behaviour of $\chi$ above $\approx$~6~K in Fig.~\ref{raw}c
together with the approximately linearly decreasing variation in $C/T$
with $T$~\cite{jaime1} at $\approx$~38~T could, nevertheless, be
possible indications of non-Fermi liquid behaviour.  Similar types of
behaviour in other $f$-electron systems have been ascribed to the
presence of a non-Fermi liquid~\cite{stewart1}.

Changes in the value of $M$ through the transitions provides clues as
to the nature of the ordered phase.  For $B\lesssim$~25~T, $M$ is
weakly dependent on temperature, exhibiting a predominantly Pauli
paramagnetic response typical of heavy Fermi
liquids~\cite{hewson1,stewart2}.  This, together with specific heat
measurements above the ordering temperature~\cite{jaime1} and
de~Haas-van~Alphen measurements below the ordering
temperature~\cite{ohkuni1,keller1}, unambiguously establishes the
existence of a heavy Fermi liquid with an {\it effective} Fermi energy
of order 10~meV. In the itinerant $f$-electron picture, a heavy Fermi
liquid results from mixing of the $f$-electrons with regular
conduction electron states~\cite{hewson1}.  When IEM occurs, the
spin-up component itinerant $f$-electron band is shifted by the Zeeman
interaction to energies just below $\varepsilon _{\mathrm{F}}$ at
$B_{\mathrm{M}}$ (see Fig.~\ref{bands}), causing $M$ to undergo a
dramatic increase by as much as 1~$\mu _{\mathrm{B}}$ per $f$-electron
atom~\cite{edwards1}.  As a result, $f$-electrons that were mostly
itinerant below $B_{\mathrm{M}}$ become mostly aligned and localised
at fields above $B_{\mathrm{M}}$.  The field $B_{\rm M}$ corresponds
to a situation where the Fermi energy $\varepsilon_{\mathrm{F}}$
intersects the middle of the spin-up $f$-electron band causing it to
be half occupied.  This leads to an approximately temperature
independent $M$ at $B_{\rm M}$ (see Fig.~\ref{raw}{\bf a}) but with
$\chi$ increasing dramatically with decreasing temparture (see
Fig.~\ref{raw}{\bf c}).  The continuation of the temperature
independence of $M$ at $B_{\rm M}$ below $\approx$~6~K, accompanied by
an abrupt reduction in $\chi$, indicates that the field-induced
ordered phase stabilises a situation where approximately half of the
$5f$-electrons become localised while the other half remain itinerant. 
This type of behaviour implies the existence of a charge gap in the
spin-up itinerant $5f$-electron band at $\varepsilon_{\mathrm{F}}$, at
$B_{\rm M}$.

The formation of a charge gap in the spin-up $5f$-electron band is
consistent with the existence of a spin-singlet order parameter that
breaks translational symmetry.  Order parameters of the charge-density
wave~\cite{gruner1} and OAF~\cite{chandra1} type both possess this
essential property; the latter also breaks time reversal symmetry. 
They both involve singlet pairing of quasiparticles at a
characteristic translational wavevector $\mathbf{Q}$, where
$\mathbf{Q}$ is determined by details of the Fermi surface
topology~\cite{chandra1,gruner1,chakravarty1}.  In fact, regardless of
the pairing symmetry, any singlet order parameter that involves
spatial variations in charge density, or relative charge densities
between one or more electron channels, will lead to such a gap.  Order
parameters of this type are also amenable to the possibility of
re-entrant behaviour (see below).  Evidence that the HO and
field-induced HO phases have a common origin may be provided by the
field and temperature dependence of $\chi$.  The transition into the
HO phase is devoid of any pronounced features in the temperature and
field dependence of $\chi$ at fields below $\approx$~34~T, while those
into the field induced phase exhibit similar behaviour over a narrow
interval between 36.8 and 37.1~T (see Fig.~\ref{contour}). 
Furthermore, all transitions into (or out of) both phases evolve into
ones that are first order in the limit $T\rightarrow$~0, although
actual magnetic hysteresis remains undetected~\cite{jaime1}.  First
order transitions give rise to pronounced maxima in $\chi$ and/or
magnetocaloric heating as the field is swept~\cite{jaime1}.  Some
degree of similarity between the low and high magnetic field phases is
also apparent in the temperature dependence of the $\chi$ in
Fig.~\ref{raw}{\bf c} at 34.0~T and 37.9~T respectively.

In order to understand how re-entrance of the HO parameter can occur,
it is instructive first to consider the density of electronic states
(DOS) within the ordered phase for $B<$~35~T, which has received most
attention thus far~\cite{chandra1,grigera1}.  Figures~\ref{bands}{\bf
a}-{\bf e} show a schematic of the evolution of the total DOS with $B$
with (black lines) and without (red lines) ordering.  At $B=$~0
(Fig.~\ref{bands}{\bf a} red line), the spin-up and and spin-down
Fermi surfaces are degenerate.  The introduction of a periodic charge
potential $V(\mathbf{r}\cdot \mathbf{Q}_{0})$ within the HO phase must
therefore result in the independent formation of band gaps for both
spins (black line).  This process is only efficient at minimising the
energy of the systems if a significant part of the DOS is gapped at
$\varepsilon_{\rm F}$.  The introduction of $B$ in
Fig.~\ref{bands}{\bf b}, however, causes the energies of the spin-up
and spin-down bands to split, leading to spin-up and spin-down Fermi
surfaces of different sizes.  The efficiency by which
$V(\mathbf{r}\cdot\mathbf{Q}_{0})$ can gap both spins therefore
becomes progressively worsened as $B$ is increased, leading to the
weakening of the gap and, eventually, to its destruction in a manner
analogous to that of reaching the Pauli limit of a singlet
superconductor~\cite{dieterich1,zanchi1,harrison1}.  A previous
magnetoresistance study has provided experimental evidence for
weakening of the gap in fields of $\approx$~25~T~\cite{mentink1}. 
Ultimately, the ordered phase must be destroyed at a critical field
$B_{\rm c}\leq B_{\mathrm{M}}$, whereupon the spin-up and spin-down
Fermi surfaces become extremely asymmetric.  The effect of $B$ on
translational symmetry-breaking spin singlet order parameters has been
extensively modeled using mean field
theory~\cite{dieterich1,zanchi1,harrison1}.  One possibility is that
the transition evolves into one that is first order that terminates at
a critical field $B_{\mathrm{c}}=\Delta _{0}/\sqrt{2}\sigma
g\mu_{\mathrm{B}}$.  Figure~\ref{bands}{\bf c} depicts the density of
electronic states at $B_{\mathrm{c}}$ where the spin-up and spin-down
$f$-electrons states have become clearly resolved and singlet gap
formation is no longer favoured.  Upon estimating the size of the
order parameter using the BCS relation $2\Delta
_{0}=3.52k_{\mathrm{B}}T_{\mathrm{o}}$~\cite{dieterich1,harrison1} and
inserting free electron parameters for the spin $\sigma=\frac{1}{2}$
and \textit{g}-factor $g=$~2, we obtain $B_{\mathrm{c}}\approx $~32~T.
This is of comparable order to the first order-like transition at
$\approx $~35~T~\cite{jaime1} obtained from the current measurements
(see Fig.~\ref{contour}) as well as specific heat~\cite{jaime1}. 
Given that the product $\sigma g\mu_{\mathrm{B}}$ in $f$-electron
systems can depart from the free electron value~\cite{hewson1}, this
agreement may be merely circumstantial.  However, a further prediction
of mean field theory is that both the transition temperature
$T_{\mathrm{o}}(B)$ as a function of $B$~\cite{dieterich1} and the
critical field $B_{\mathrm{c}}(T)$ as a function of
$T$~\cite{harrison1} intersect the axes in a perpendicular manner, and
both can be expanded in series of even powers of $B$ and $T$
respectively.  A plot of $T_{\mathrm{o}}^2$ versus $B^2$ should
therefore yield a line that intercepts both axes in an approximately
linear fashion.  This is confirmed in Fig.~~\ref{bands}{\bf f} upon
making such a plot with actual URu$_2$Si$_2$ data.  An interesting
situation then develops in the vicinity of $B_{\mathrm{M}}$, enabling
the realisation of a re-entrant hidden order (RHO) phase with a
modified translational vector $\mathbf{Q}_\downarrow$.  Strong
magnetic fluctuations at $B_{\rm M}$ can be associated with the
vanishingly small energy that separates spin-up electrons in localised
and itinerant states at $\varepsilon_{\rm F}$~\cite{millis1}.  This,
combined with the weak dispersion of the spin-up $f$-electron band,
makes the system especially vulnerable to forming an ordered phase. 
Ordering is especially easy to realise if the periodic potential
$V(\mathbf{r}\cdot\mathbf{Q}_\downarrow)$ becomes comparable to the
bandwidth of the spin-up $f$-electrons, because it will succeed at
gapping the entire spin-up density of $f$-electron states regardless
of the value of $\mathbf{Q}_\downarrow$ and regardless of the absence
of well defined spin-up momentum quantum numbers.  A significant
amount of energy is gained by opening such a gap at $\varepsilon_{\rm
F}$, and this would then account for the observed narrow gap in the
spin-up $f$-electron band (see Fig.~\ref{bands}d).  The value of
$\mathbf{Q}_\downarrow$ need only be optimised to match the topology
of the spin-down Fermi surface, which continues to be present at
$B_{\mathrm{M}}$.  Once $B>B_{\mathrm{M}}$, ordinary Fermi liquid
behaviour is expected to be restored, but with the spin-up
$f$-electrons being fully polarized as depicted in
Fig.~\ref{bands}{\bf e}.  The pronounced asymmetry between spins in
the Fermi liquid makes the formation of an ordered phase unlikely,
enabling the emergence of a Schottky anomaly in the specific
heat~\cite{jaime1}.

In summary, we present $M$ data which shows that the HO parameter is
first destroyed by Zeeman splitting in a magnetic field but then
restored in a re-entrant phase~\cite{jaime1}.  The $T$ and $B$
dependence of $\chi$ reveals that IEM plays a role in its re-entrance,
possibly indicating that HO is restored in response to magnetic
fluctuations in the vicinity of a metamagnetic quantum critical end
point~\cite{grigera1,millis1}.  If true, this could be the first
observation of the creation of an ordered phase in the vicinity of a
magnetic field-induced quantum critical end point.  We propose the
existence of separate HO and RHO phases characterised by a common
spin-singlet translational symmetry-breaking order parameter with
slightly different translational vectors $\mathbf{Q}_0$ and
$\mathbf{Q}_\downarrow$.

This work is supported by the National Science Foundation, the
Department of Energy and Florida State.  We thank Christian Batista,
Kee-Hoon Kim, Greg Boebinger and John Singleton for useful
discussions.

\begin{figure}
\caption{The $B>$~30~T versus $T$ phase diagram of
URu$_2$Si$_2$ combined with a colour intensity plot of $\chi$ measured
at many different temperatures.  Square, triangle and circle symbols
mark $B_{\rm M}$ and transitions into and out of the HO and RHO hidden
order (RHO) phases respectively.  The curved dotted lines depict the
continuation of the phase boundaries revealed by specific heat and
transport studies~\cite{jaime1}.}
\label{contour}
\end{figure}

\begin{figure}
\caption{{\bf a}, $M$ of URu$_2$Si$_2$ at several different
temperatures for $B$ applied along the $c$-axis.  {\bf b}, $\chi$ in
the vicinity of $B_{\rm M}$ at several temperatures above the
re-entrant ordering temperature.  {\bf c} $\chi$ at $B_{\rm M}$ and at
$B\approx$~34~T as a function of $T$.}
\label{raw}
\end{figure}

\begin{figure}
\caption{{\bf a}-{\bf e} Schematics of the evolution of the total DOS
in URu$_2$Si$_2$ with $B$ (as indicated) before (red lines) and after
(black lines) formation of the HO or RHO phases.  Prior to ordering,
mixing between conduction electron states and $f$-electron states
gives rise to a large ``Abrikosov-Suhl resonance''-like
feature~\cite{hewson1}.  {\bf f} A plot of the transition temperature
squared $T^2_{\rm o}$ versus the magnetic field squared $B^2$ taken
from specific heat and transport data in Reference~\cite{jaime1}.}
\label{bands}
\end{figure}

\end{document}